\documentclass[9pt,twocolumn,twoside]{pnas-new}
\usepackage{graphicx}  
\usepackage{dcolumn}   
\usepackage{bm}        
\usepackage{amssymb}   
\usepackage{amsmath}
\usepackage{float}
\usepackage{xcolor}

\templatetype{pnasresearcharticle} 

\title{Is There a Scaling Law in the Inviscid Coalescence of Unequal-size Droplets?}

\author[a,1]{Xi Xia}
\author[b,c,1]{Yicheng Chi} 
\author[b,2]{Peng Zhang}

\affil[a]{School of Mechanical Engineering, Shanghai Jiao Tong University, Shanghai 200240, P. R. China}
\affil[b]{Department of Mechanical Engineering, City University of Hong Kong, Kowloon Tong 999077, Hong Kong}
\affil[c]{School of Automotive and Transportation Engineering, Shenzhen Polytechnic University, Shenzhen, Guangdong 518055, P. R. China}

\leadauthor{Xia} 

\significancestatement{This work fills in the gap of liquid bridge evolution during droplet coalescence in the inviscid regime when size disparity exists between the two droplets. Compared with the classical 1/2 power-law scaling for the equal-size case, our experimental data exhibit notably increased prefactor and deviation from the power-law scaling when the size ratio enlarges. To resolve the size-disparity effect, we develop the first theoretical model for the asymmetric liquid bridge evolution, yielding the collapse of different-size-ratio data onto a single curve. More importantly, this model suggests an exponential dependence of the bridge radius on time, which exhibits a scaling-free feature at late coalescence time and large size ratio, further justifying the observed breaking of the scaling law.}

\authorcontributions{Author contributions: X.X. and P.Z. developed the theory; Y.C. and P.Z. conducted the experiment; X.X. and Y.C. performed the analysis; and X.X. and P.Z. wrote the paper.}
\authordeclaration{The authors declare no conflict of interest.}
\equalauthors{\textsuperscript{1}X.X. and Y.C. contributed equally to this work.}
\correspondingauthor{\textsuperscript{2}To whom correspondence should be addressed. E-mail: penzhang@cityu.edu.hk}


\begin{abstract}
This work examines the coalescence of two unequal-size spherical liquid droplets in the inviscid regime, with an emphasis on exploring the scaling of the liquid bridge evolution. Our experiment suggests that the classical 1/2 power-law scaling for equal-size droplets still holds for the unequal-size situation of small size ratios, but it diverges as the size ratio increases. Employing an energy balance analysis, we develop a theoretical model to collapse the experimental data of different droplet size ratios. The model reveals an exponential dependence of the bridge's radial growth on time, implying an intrinsic breaking of scaling law. The scale-free evolution behavior is evident only at late coalescence time and large size ratio, which can be explained using the length and time scales obtained from the theory.     
\end{abstract}

\dates{This manuscript was compiled on \today}
\doi{\url{www.pnas.org/cgi/doi/10.1073/pnas.XXXXXXXXXX}}

\begin{document}

\maketitle
\thispagestyle{firststyle}
\ifthenelse{\boolean{shortarticle}}{\ifthenelse{\boolean{singlecolumn}}{\abscontentformatted}{\abscontent}}{}

\dropcap{D}uring natural and industrial processes, the contact or impact of droplets \cite{EggersJ:99a,AartsDGAL:05a,YarinAL:06a,ThoravalMJ:12a,TranT:13a} could lead to the coalescence of liquid-gas interfaces, which is crucial to the outcome or performance of the relevant applications. Extensive research has been carried out to understand the most basic situation, the momentumless coalescence of a pair of liquid droplets \cite{Thoroddsen:05a,PaulsenJD:11a,ZhangP:11a,KavehpourHP:15a,EggersJ:25a}. Early studies \cite{EggersJ:99a,DucheminL:03a} found that the radial growth of the liquid bridge, which forms between the merging droplets, satisfies certain scaling relations between the radius $R$ of the bridge and time $t$. Later experimental \cite{WuM:04a,AartsDGAL:05a,Thoroddsen:05a,FezzaaK:08a,CaseSC:09a} and numerical \cite{BurtonJC:07a,PothierJC:12a,GrossM:13a,SprittlesJE:12a} works were able to confirm the existence of a $1/2$ power-law scaling when the coalescence is in the inviscid (or inertial) regime. On the other hand, the bridge evolution in the viscous regime is better modeled by a linear scaling \cite{AartsDGAL:05a,Thoroddsen:05a,YaoW:05a,BurtonJC:07a,PaulsenJD:11a}. The crossover (or transition) \cite{BurtonJC:07a} between the viscous and inertial regimes has also attracted considerable interest, from the discovery of a master curve for both regimes \cite{PaulsenJD:11a,PaulsenJD:13a} to the development of theoretical models justifying the underlying universality \cite{Xia:19b,HackMA:20a}.

Previous research on binary droplet coalescence mainly revolves around two equal-size droplets. However, less attention has been given to droplet coalescence with size disparities, despite its higher relevance to reality. Among the existing works involving the coalescence of unequal-size droplets \cite{AnilkumarAV:91a,BlanchetteF:10a,ZhangP:13a,ZhangP:15a,ZhangP:16a,Xia:17a}, the main focus was on the effect of internal mixing facilitated by the breaking of symmetry. Regarding the evolution of the liquid bridge, it is evident from our previous work \cite{Xia:17a} that the bridge surface shows an asymmetric growth--the bridge interface becomes tilted as it expands out. Yet, little quantitative study exits on the liquid bridge evolution of unequal-size droplet coalescence. Especially, an intriguing question is whether the $1/2$ power-law scaling still exists in the asymmetric coalescence scenario.  

\section*{Experiment} 
    
\begin{figure}
\begin{center}
\includegraphics[scale=0.34]{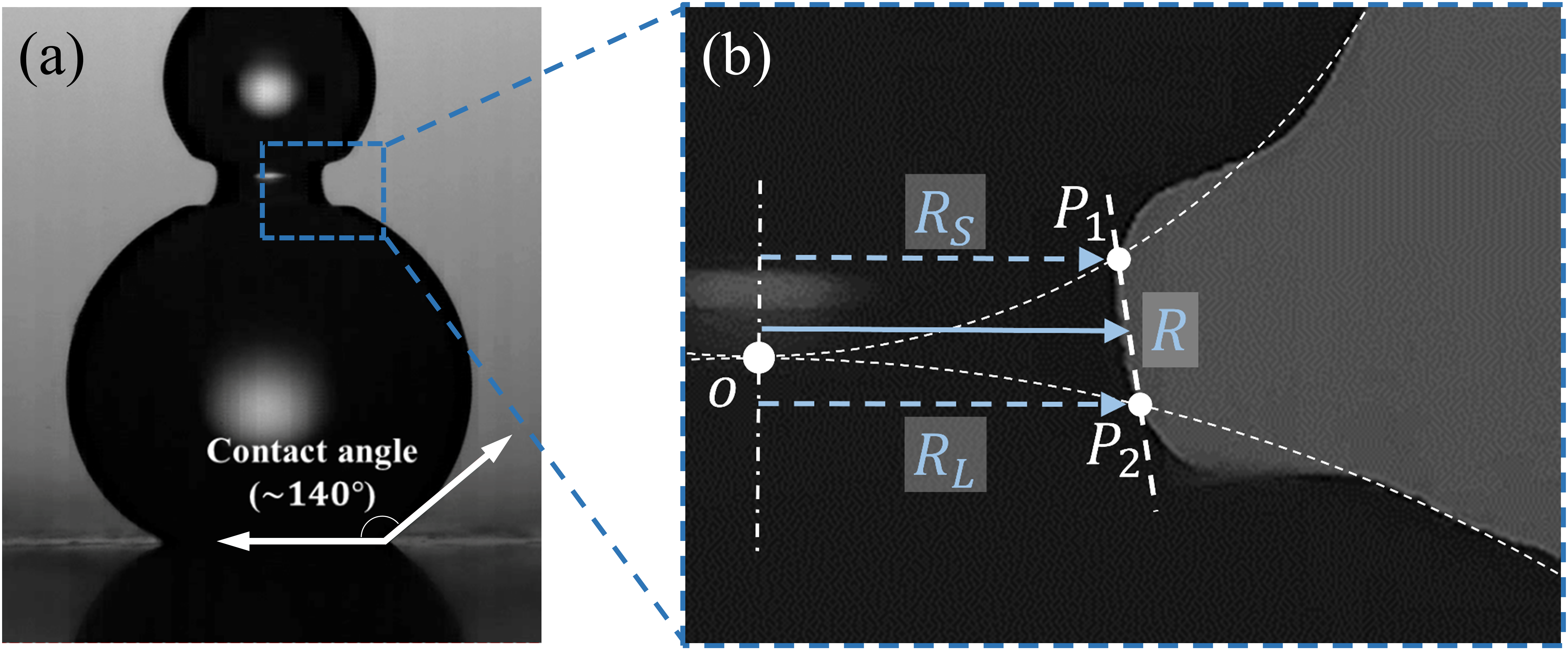}
\caption{(a) high-speed image of the sessile and pendant droplets and (b) zoomed-in detail of the liquid bridge.}
\vspace{-6mm}
\label{fig:setup}
\vspace{-2mm}
\end{center}
\end{figure}

Regarding the coalescence of two unequal-size droplets, we first conduct experiment to resolve the bridge's temporal evolution for droplet pairs of various liquid properties and size ratios. Our experiment adopts the sessile-pendant approach for droplet coalescence, which is detailed in \emph{Materials and Methods}. Fig.~\ref{fig:setup}(a) shows a snapshot of two unequal-size merging droplets. The size disparity is quantified by the droplet size ratio, $\Delta = D_L/D_S$, where $D_S$ and $D_L$ are the diameters of the small and large droplets, respectively. Fig.~\ref{fig:setup}(b) illustrates the zoomed-in detail of the liquid bridge, exhibiting a distinct asymmetric bridge interface. As such, we can define two characteristic radii of the droplet bridge, $R_S$ and $R_L$, respectively as the radial distances from the two points, $P_1$ and $P_2$, where the bridge interface intersects the contours of the initial droplets (denoted by the white-dashed lines), to the axis of symmetry. Then, the characteristic radius of the circular bridge, $R$, is defined as $R = (R_S + R_L)/2$. Note this definition differs from the equal-size situation, where the bridge radius is defined as the minimum radial distance of the bridge interface to the center axis; thus, the unequal-size coalescence yields a slightly larger $R$ by definition.

\begin{figure}
\begin{center}
\includegraphics[scale=0.38]{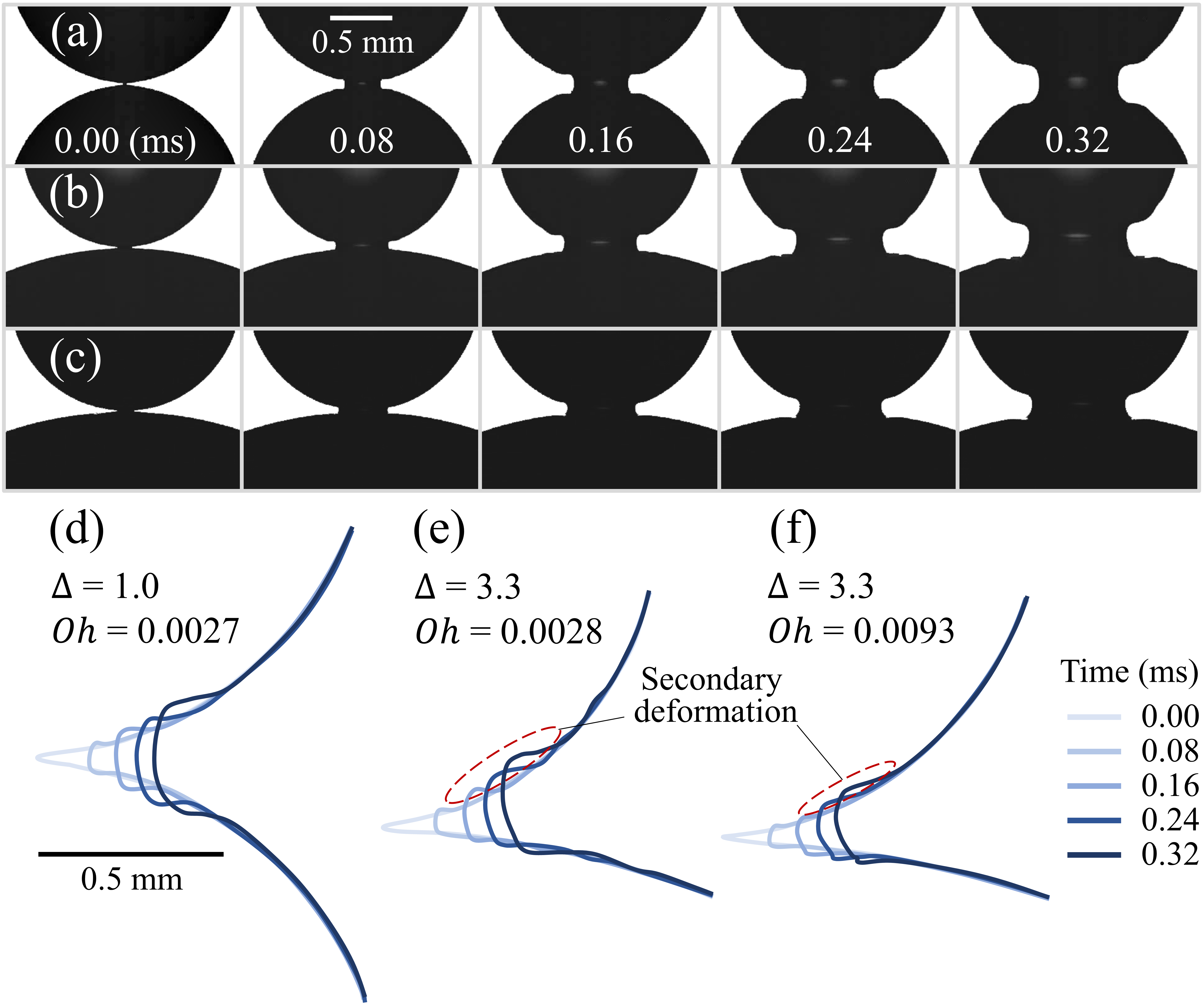}
\caption{Image sequences of the coalescence process of (a) equal-size water droplets and (b-c) unequal-size droplets of water and 40 wt$\%$ aqueous glycerol. (d-f) show the evolutions of the extracted interface contours corresponding to (a-c).}
\vspace{-6mm}
\label{fig:exp}
\vspace{-2mm}
\end{center}
\end{figure}

To study the effect of varying liquid properties, i.e., density $\rho_l$, dynamic viscosity $\mu_l$, and surface tension $\sigma$, we adopt water and two aqueous glycerol solutions with 40 wt$\%$ and 60 wt$\%$ glycerol. These different liquids can be characterized by the non-dimensional Ohnesorge number, which is defined as $Oh = \mu_l (\rho_l \sigma D_S)^{-1/2}$. Based on the liquid properties specified in \emph{Materials and Methods}, $Oh$ numbers of all droplets in this experiment are estimated to vary among $10^{-3}$-$10^{-2}$, which belongs to the inviscid or inertia coalescence regime \cite{Xia:19b}. The image sequences for representative droplet coalescence cases are presented in Fig.~\ref{fig:exp}(a-c), where the droplet interface contours corresponding to the different snapshots are extracted and overlapped in Fig.~\ref{fig:exp}(b-f). We observe that the liquid bridge displays a notable asymmetry for $\Delta > 1$, rendering an inclined bridge interface with $R_L > R_S$, which becomes increasingly evident as the bridge evolves with time. As $Oh$ increases from 0.0028 to 0.0093, the bridge profiles in Fig.~\ref{fig:exp}(f) follow the original droplet contour more closely than those in Fig.~\ref{fig:exp}(e), as the secondary deformation on the bridge's upper and lower surfaces tends to be inhibited. This can be interpreted that the primary bridge movement gives rise to the development of a capillary wave along the droplet surface, which can be damped by enhanced viscosity.   

\begin{figure}
\begin{center}
\includegraphics[scale=0.48]{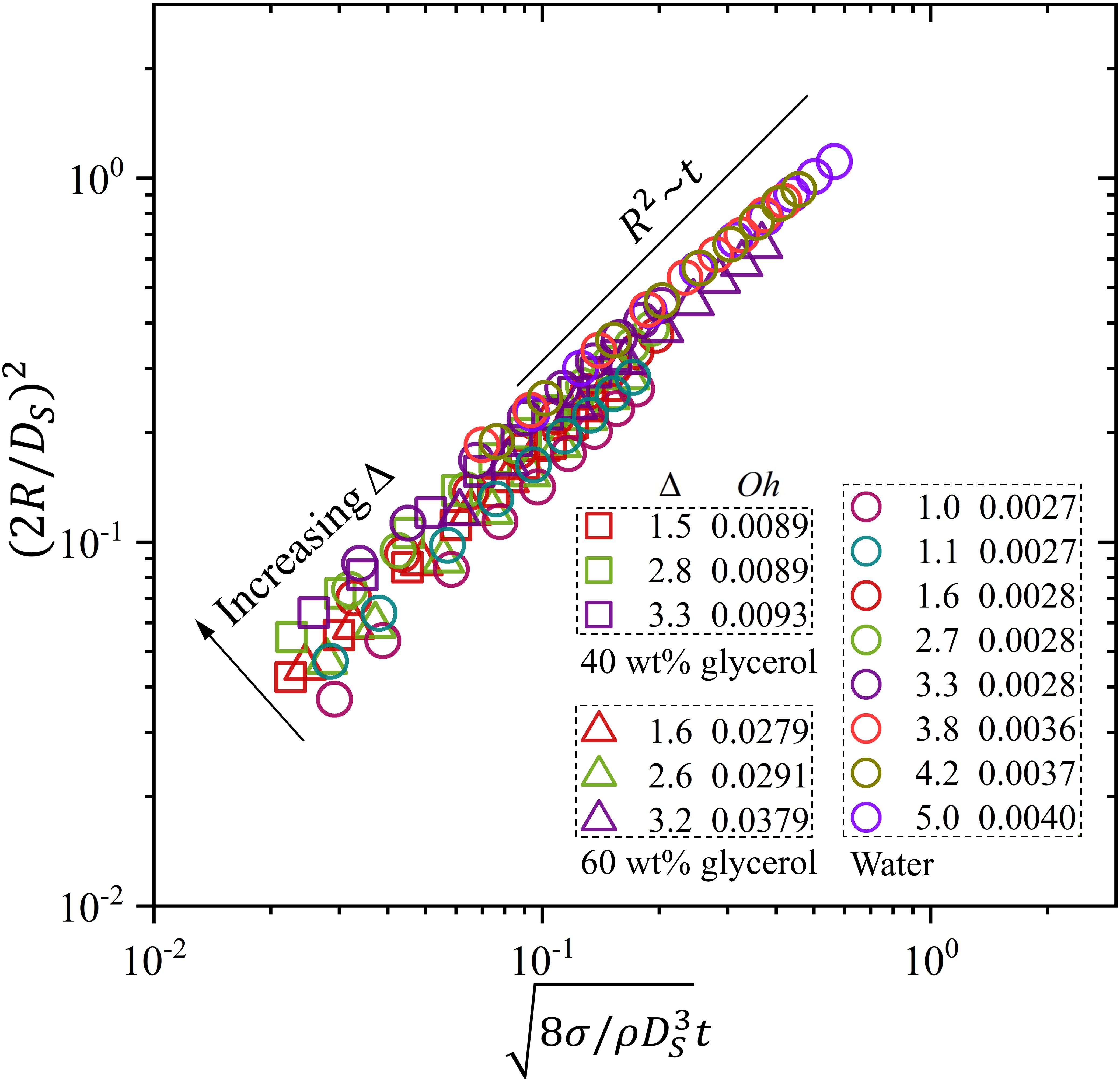}
\vspace{-2mm}
\caption{Effect of $\Delta$ on the scaling of $R^2 \sim t$. See Supplemental Materials \cite{SM:24a} for detailed experimental parameters.}
\label{fig:4}
\vspace{-5mm}
\end{center}
\end{figure}

Based on the evolution of the liquid bridge interface extracted from the experimental results of various $\Delta$ and $Oh$, we now check whether the classical scaling law of $R \sim t^{1/2}$ holds for unequal-size droplet coalescence, by plotting the data in the parameter space $[(8\sigma/\rho D_S^3)^{1/2}t, (2R/D_S)^2]$ \cite{DucheminL:03a, AartsDGAL:05a} in Fig.~\ref{fig:4}. While the overall scaling of $R^2 \sim t$ is still valid for most cases, there is an apparent upward drift of data from the $\Delta = 1.0$ line as $\Delta$ increases. This can be understood that the presence of a larger droplet enhances the expansion speed of the bridge interface, corresponding to a larger prefactor of the power-law scaling. Furthermore, the cases with $\Delta > 3$ display slightly decreased slopes compared with that of $R^2 \sim t$ as time proceeds to the later stage of coalescence when the liquid bridge becomes more asymmetric, indicating a mild diversion from the classical power-law scaling. 

\section*{Theory}

To further understand the different scaling characteristics associated with unequal-size droplet coalescence, we next develop a theory to model the radial evolution of the liquid bridge in the inertial regime. Inspired by the liquid bridge configuration observed from the experiment, we present a geometric model for the bridge surface between two unequal-size droplets in Fig.~\ref{fig:1}, based on the core geometric variables of $R_S(t)$, $R_L(t)$, $\theta_S(t)$, $\theta_L(t)$, and $S(t)$. 

\subsection*{Assumptions}
In this model, the geometric correlations are derived based on the following assumptions.\\
(i) `Similar-size droplets', meaning the two droplets are of similar sizes such that $\Delta \sim O(1)$, $R_L/R_S \sim O(1)$, and $\theta_L/\theta_S \sim O(1)$.\\ 
(ii) `Small bridge', meaning that the characteristic radii of the bridge, $R_S$ and $R_L$, are much smaller than the droplet diameters, i.e., $R_S/D_S \sim o(1)$ and $R_L/D_L \sim o(1)$. This assumption is readily satisfied during the early-stage coalescence.\\ 
(iii) `Equally-dividable bridge interface', meaning there exists a principle normal direction $\mathbf{n}_p$ dividing the bridge interface into similar-shaped upper and lower sections, and the surface stress (including both pressure difference $\Delta p$ and viscous stress $\mathbf{\tau}$) is also equally distributed over the two sections. In other words, the asymmetry of the bridge only manifests in its tilted normal direction, while its shape remains almost symmetric. Considering the present experimental results, this assumption can be considered a first approximation under the condition of (ii), although the two sections divided by $\mathbf{n}_p$ might not have perfectly equal shape.\\

\subsection*{Geometry}
\begin{figure}
\begin{center}
\includegraphics[scale=0.99]{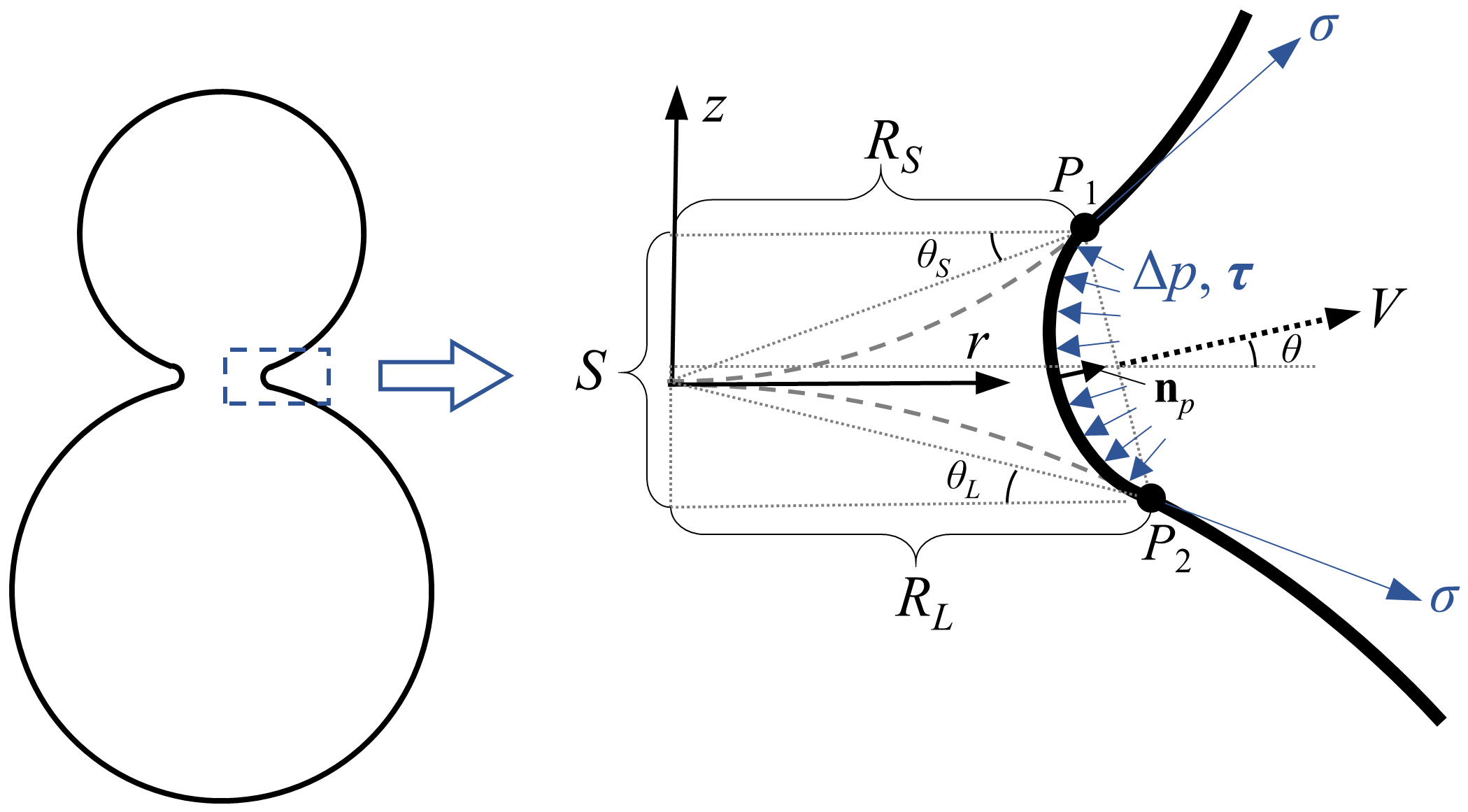}
\caption{Schematic of the liquid bridge between two unequal-size merging droplets.}
\vspace{-5mm}
\label{fig:1}
\vspace{-3mm}
\end{center}
\end{figure}

In the model illustrated in Fig.~\ref{fig:1}, we have the following geometric correlations, 
\begin{equation}
\begin{split}
\label{eq:m1}
&\frac{R_S}{D_S} = \frac{\sin{(2\theta_S)}}{2} = \theta_S - O(\theta_S^3),\\ 
&\frac{R_L}{D_L} = \frac{\sin{(2\theta_L)}}{2} = \theta_L - O(\theta_L^3).
\end{split}
\end{equation}
Applying assumption (ii), it is deduced that both $\theta_S$ and $\theta_L$ are of $o(1)$, so the high-order terms in Eq.~\ref{eq:m1} can be neglected. It follows that the width of the bridge, $S$, defined as the axial distance between $P_1$ and $P_2$, satisfies
\begin{equation}
\begin{split}
\label{eq:m2}
S &= R_S \tan{\theta_S} + R_L \tan{\theta_L}\\
  &= R_S\theta_S + R_L\theta_L + R_S O(\theta_S^3) + R_L O(\theta_L^3).
\end{split}
\end{equation}
Combining Eqs.~\ref{eq:m1} and~\ref{eq:m2}, we can easily show that $S/D_S$ and $S/D_L$ are of $O(\theta_S^2)$ or $O(\theta_L^2)$, respectively.

Note that, compared with the equal-size situation, the unequal-size coalescence is characterized by the tilted bridge interface and the misaligned bridge movement from the radial direction, as delineated in Fig.~\ref{fig:1}. The cause of the tilted interface may be understood from a force balance analysis. Under assumption (iii), the principle normal direction $\mathbf{n}_p$ of the bridge interface should align with the integral of the total surface stress across the interface, which is balanced by the surface tension exerted at both ends of the bridge interface. Given the two surface tension forces are identical to each other, $\mathbf{n}_p$ must be in line with the angular bisector of the two surface tension forces. Therefore, the tilting angle of the bridge interface, $\theta$, can be simply expressed as 
\begin{equation}
\label{eq:m3}
\theta = \theta_S - \theta_L.
\end{equation}
Since both $\theta_S$ and $\theta_L$ are of $\sim o(1)$, Eq.~\ref{eq:m3} implies that $\theta$ is also a $o(1)$ quantity. Thus, $\theta$ is treated as the `small parameter' hereinafter. From Fig.~\ref{fig:1}, we further note another geometric relationship, $R_L-R_S = S\tan{\theta} \approx S \theta$, which can be combined with Eqs.~\ref{eq:m1} and~\ref{eq:m3} to derive
\begin{equation}
\begin{split}
\label{eq:m5}
\theta &\approx \left(\frac{1}{D_S} - \frac{1}{D_L} \right)\frac{R_L + R_S}{2} - \left(\frac{1}{D_S} + \frac{1}{D_L} \right)\frac{R_L - R_S}{2}\\
       &\approx \left(\frac{1}{D_S} - \frac{1}{D_L} \right)R - \left(\frac{1}{D_S} + \frac{1}{D_L} \right)\frac{S \theta}{2}\\
			 &\approx \left(\frac{1}{D_S} - \frac{1}{D_L} \right)R + O(\theta_S^2) \theta,
\end{split}
\end{equation}
where the second term on the right-hand side is negligible; its derivation is based on $S/D_S \sim O(\theta_S^2)$ and $S/D_L \sim O(\theta_L^2)$ that follow from Eq.~\ref{eq:m2}. Similarly, we can apply $R_L-R_S \approx S \theta$ to Eqs.~\ref{eq:m1} and~\ref{eq:m2} to obtain
\begin{equation}
\begin{split}
\label{eq:m6}
S &\approx \left(\frac{1}{D_S} + \frac{1}{D_L} \right)R^2 + \left(\frac{R_L+R}{D_L} - \frac{R_S+R}{D_S} \right)\frac{S \theta}{2}\\
			 &\approx \left(\frac{1}{D_S} + \frac{1}{D_L} \right)R^2 + O(\theta^2) S,
\end{split}
\end{equation}
where the second term on the right-hand side can also be dropped off.

\subsection*{Kinematics}
According to assumption (iii), the viscous stress expressed as $\mathbf{\tau} = 2\mu\mathbf{S}$ (with $\mathbf{S}$ being the strain-rate tensor) is also equally distributed between the upper and lower sections of the liquid bridge interface. It follows that the liquid-side velocity must also be equally distributed over the bridge interface, so the overall bridge movement is inline with its principle normal direction $\mathbf{n}_p$. The physical implication here is significant. Consider the very early stage of coalescence when $R_S \approx R_L$, Eq.~\ref{eq:m3} dictates that $\theta$ is a positive value, which explains why the bridge movement is inclined towards the smaller droplet from the beginning. It follows that the velocity $V$ at which the bridge interface expands out is given by the kinematic relationship, $\mathrm{d} R/\mathrm{d} t = V \cos{\theta}$. Applying Taylor expansion, it takes the form
\begin{equation}
\label{eq:m4}
\frac{\mathrm{d} R}{\mathrm{d} t} \approx V \left(1-\frac{\theta^2}{2} \right).
\end{equation}
Note that Eq.~\ref{eq:m4} recovers the equal-size coalescence kinematics with vanishing $\theta$. 

\subsection*{Energy balance}
Now, considering the energy balance during the coalescence process, the movement of the liquid entrained by the bridge is driven by the rapid discharge of the surface energy, which can be expressed as
\begin{equation}
\label{eq:m7}
\Delta E_s + \Delta E_k \approx 0,
\end{equation}
where $\Delta E_s$ and $\Delta E_k$ represent the respective changes in surface and kinetic energies from the initial state prior to coalescence. Note Eq.~\ref{eq:m7} requires the viscous dissipation being negligible compared to the change in inertia energy. Based on the geometries in Fig.~\ref{fig:1}, we have
\begin{equation}
\begin{split}
\label{eq:m8}
\Delta E_s &= -\pi D_S^2 (\theta_S^2+O(\theta_S^4)) \sigma - \pi D_L^2 (\theta_L^2+O(\theta_L^4)) \sigma\\
  &+ \pi S (R_S+R_L)[1 + O((\theta_S+\theta_L)^2)] \sigma,
\end{split}
\end{equation}
where $\sigma$ is the surface tension. Since $\theta_S$ and $\theta_L \sim o(1)$, we can combine Eqs.~\ref{eq:m1} and~\ref{eq:m2} to obtain $S \approx R_S\theta_S + R_L\theta_L \approx D_S\theta_S^2 + D_L\theta_L^2$. Thus, the third term on the right-hand side of Eq.~\ref{eq:m8} has the leading order of $O(\theta_S^3)D_S^2\sigma$ or $O(\theta_L^3)D_L^2\sigma$, which is one order smaller than the first two terms and can be dropped off. Next, $\Delta E_k$ can be estimated as
\begin{equation}
\label{eq:m9}
\Delta E_k \approx \frac{C}{2} \pi \rho_l R^2S V^2,
\end{equation}
where $\rho_l$ is the liquid density and $C$ is a prefactor related to the volume and velocity distribution of the moving fluid entrained with the bridge. Note that the leading-order energy terms in Eqs.~\ref{eq:m8} and~\ref{eq:m9} are almost identical to those of the equal-size situation, except for the parameters associated with the different droplet sizes. 

With $S$ given by Eq.~\ref{eq:m6}, we can plug Eqs.~\ref{eq:m8} and~\ref{eq:m9} into Eq.~\ref{eq:m7} and balance the leading order terms to obtain
\begin{equation}
\label{eq:m10}
\frac{C}{2} \rho_l R^4 \left(\frac{1}{D_S} + \frac{1}{D_L} \right) V^2 \approx 2R^2 \sigma.
\end{equation}
Further substituting $V$ using Eqs.~\ref{eq:m4} and~\ref{eq:m5}, we have
\begin{equation}
\label{eq:m11}
\frac{1}{1-\beta R^2/2} \frac{\mathrm{d} R}{\mathrm{d} t} \approx \frac{\gamma}{R}.
\end{equation}
In simplifying Eq.~\ref{eq:m11}, we define $\beta$ and $\gamma$ as
\begin{equation}
\begin{split}
\label{eq:m12}
\beta = \left(\frac{1}{D_S} - \frac{1}{D_L} \right)^2 \text{ and }
\gamma = \left[\frac{4 \sigma D_S}{C \rho_l (1+1/\Delta)} \right]^{1/2}.
\end{split}
\end{equation}
Eq.~\ref{eq:m11} has the solution:
\begin{equation}
\label{eq:m13}
R^2 \approx \frac{2 - 2e^{-\beta \gamma t}}{\beta},
\end{equation}
where the initial condition, $R(t=0) = 0$, is used in the derivation. Eq.~\ref{eq:m13} can further take the nondimensional form of
\begin{equation}
\label{eq:m14}
R^{*2} \approx 2 - 2e^{-t^*},
\end{equation}
where $R^* = R \beta^{1/2}$ and $t^* = t\beta\gamma$. Interestingly, the bridge evolution governed by Eq.~\ref{eq:m13} or~\ref{eq:m14} no longer has a power-law scaling between $R$ and $t$.

\section*{Results and Discussion}

It is worth discussing the features of this scaling-free solution. Mathematically, the exponential dependence of $R$ on $t$ arises from the $\theta^2$ term in Eq.~\ref{eq:m4}. If $\theta = 0$ is set in Eq.~\ref{eq:m4}, the above analysis would give the exact scaling relation of $R \sim t^{1/2}$. In this sense, the breaking of the power-law scaling results from the misaligned movement of the liquid bridge from the radial direction, which changes the dependence of $\mathrm{d} R/\mathrm{d} t$ on $R$ from the -1 power to the +1 power. Furthermore, in Eq.~\ref{eq:m14}, by letting $\Delta \rightarrow 1$, we have $\beta \rightarrow 0$ and $R^2 \approx 2[1 - (1-\beta \gamma t)]/\beta = 2\gamma t$. This means that the present model is able to recover the inviscid scaling law of $R \sim t^{1/2}$ in the equal-size limit. Likewise, we attain $R^{*2} \approx 2t^*$ when $t^* \rightarrow 0$, suggesting that in practice the power-law scaling still holds for the early-stage coalescence of unequal-size droplets. This is because the misalignment effect of the liquid bridge is negligible during the initial evolution process.     

\begin{figure}
\begin{center}
\includegraphics[scale=0.46]{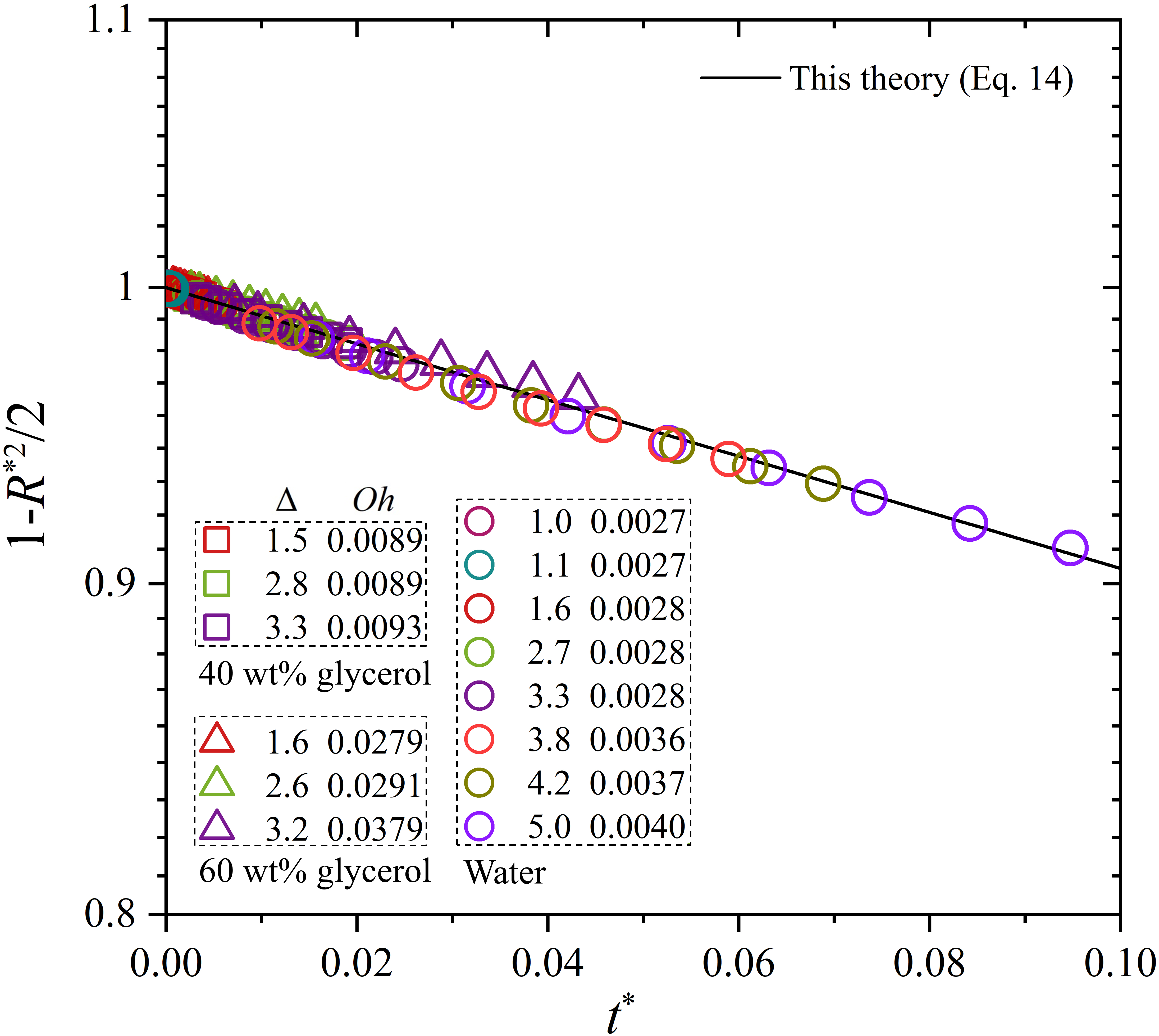}
\vspace{-2mm}
\caption{Validation of Eq.~\ref{eq:m14} against experimental data in semi-logarithmic coordinate.}
\label{fig:5}
\vspace{-5mm}
\end{center}
\end{figure}

To evaluate our theory, Fig.~\ref{fig:5} shows the validation of Eq.~\ref{eq:m14} based on the same set of data in Fig.~\ref{fig:4}. We can observe the collapse of non-unity-$\Delta$ data onto a single line given by Eq.~\ref{eq:m14} with $C = 6$ ($C$ is obtained by fitting Eq. 14), confirming the validity of the theory for droplet coalescence of various size ratios in the inviscid regime characterized by small $Oh$. And the late-stage asymmetric coalescence dynamics is successfully predicted for large size-ratio cases. This theory also determines the characteristic length and time scales of unequal-size droplet coalescence to be $\beta^{-1/2}$ and $(\beta\gamma)^{-1}$, respectively, both of which depend on the size ratio $\Delta$. Accordingly, a larger $\Delta$ generally corresponds to larger $R^*$. 

\begin{figure}
\begin{center}
\includegraphics[scale=0.44]{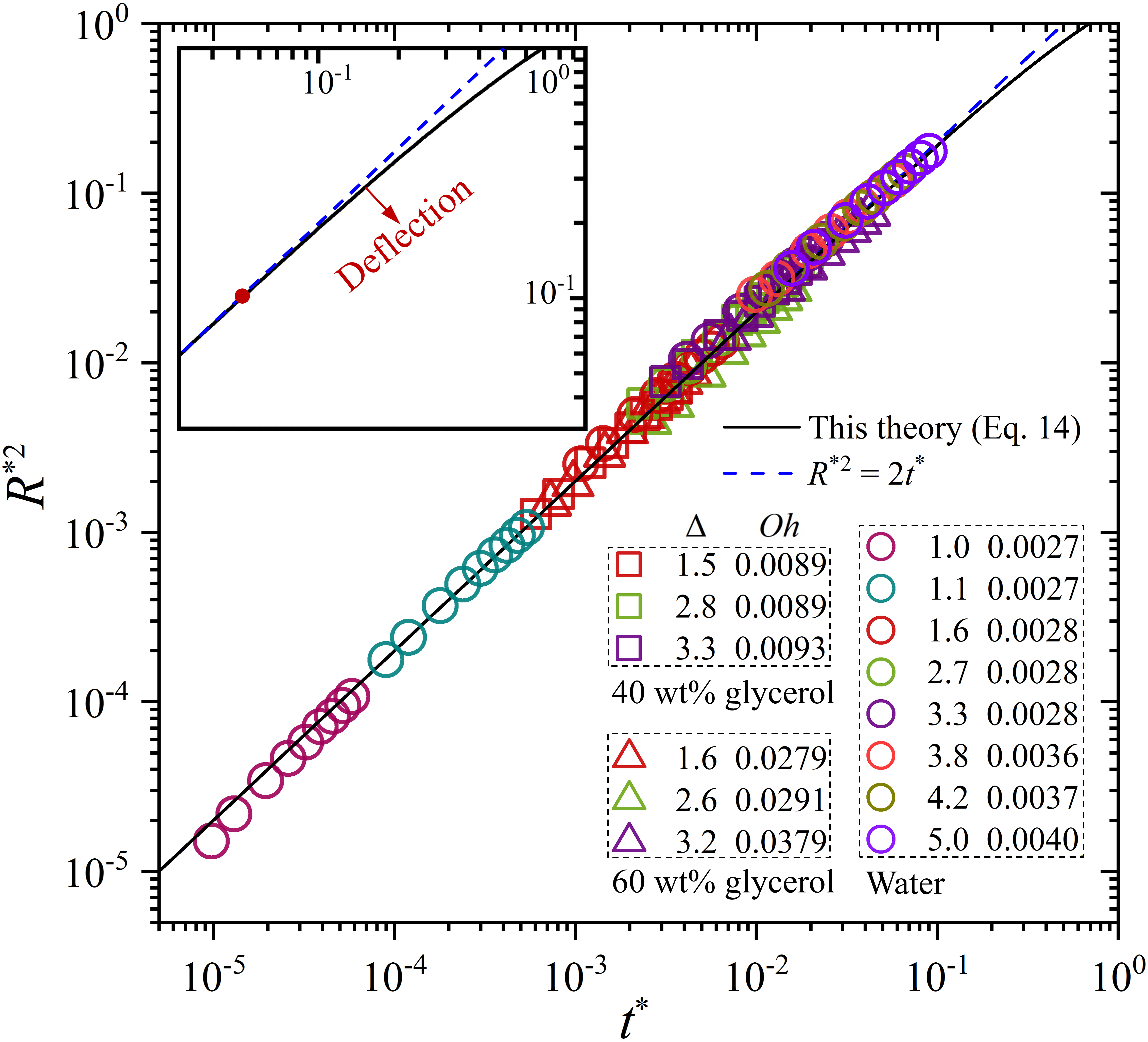}
\vspace{-2mm}
\caption{Comparison of this theory (Eq.~\ref{eq:m14}) with the power-law scaling.}
\label{fig:6}
\vspace{-5mm}
\end{center}
\end{figure}

To further illustrate to what extent the unequal-size droplet coalescence deviates from the power-law scaling, Fig.~\ref{fig:6} present the comparison of data with Eq.~\ref{eq:m14} in the $R^{*2}-t^*$ diagram. Again, the theory shows a good agreement with the experimental data. The theoretical line here remains almost linear when $R^{*2}<10^{-1}$, while it exhibits a slight deflection or deviation from the power-law scaling of $R^{*2} = 2t^*$ when $R^{*2}>10^{-1}$. To understand the corresponding $R$ range, we can further express $R^{*}$ as $(D_L-D_S)D_L^{-1} R/D_S$, which is generally much smaller than 0.3 for small-$\Delta$ cases because $(D_L-D_S)D_L^{-1}$ is rather small. As $\Delta$ increases, $(D_L-D_S)D_L^{-1}$ gradually approaches unity and the scaling line deflects as $R$ reaches the magnitude of approximately $0.3D_S$. This explains our observation from Fig.~\ref{fig:4} that the violation of the scaling becomes observable only during a rather late stage of coalescence and when $\Delta$ is greater than 3 or so. At this point, two key messages can be inferred about the coalescence of unequal size droplets. On the one hand, our theoretical mode implies an intrinsically scaling-free nature in the liquid bridge evolution process. On the other hand, most coalescence cases in practice still exhibit the classical $1/2$ power-law scaling because, according to our theory, notable scaling-breaking behavior only appears during the late-stage coalescence of large-$\Delta$ cases.

Last, the viscous effect is briefly discussed. Based on our previous theory \cite{Xia:19b}, the viscous regime for equal-size droplet coalescence occurs for $R/(Oh D_S) < 1$. In the present work, as $Oh$ varies from $10^{-3}$ to $10^{-2}$, the viscous regime corresponds to $R/D_S$ being smaller than $10^{-3}$-$10^{-2}$, for which the evolution process is beyond the resolution of this experiment. Since this range is orders-of-magnitude smaller than 0.3 (the critical radius showing apparent scaling break), it can be speculated that the viscous regime of non-equal-size droplet coalescence does not display notable difference from the equal-size situation. However, if $Oh$ increases to be $O(10^{-1})$ or higher, the viscous term in the energy balance equation (Eq.~\ref{eq:m7}) might not be ignored so that the viscous scaling could also change.

To summarize, this study experimentally and theoretically examines the scaling law for the liquid bridge evolution during the coalescence of unequal-size droplets. By accounting for the asymmetric motion of the liquid bridge interface, we attain the first theoretical solution for the inviscid bridge evolution based on balancing the changes in surface and kinetic energies. The model well captures the droplet coalescence dynamics of various size ratios by resolving the fundamental length and time scales. Although the solution is mathematically scaling-free, the emergence of scaling-breaking behaviors in reality depends on the liquid bridge being greatly asymmetric, which is satisfied only at sufficiently late coalescence time for large size-ratio cases.  

\matmethods{
\subsection*{Experimental approach}
The present experiment employs the classical sessile-pendant approach for droplet coalescence, similar to those reported previously \cite{Thoroddsen:05a,AartsDGAL:05a,FezzaaK:08a,CaseSC:09a}. A schematic of the setup is illustrated in Fig.~\ref{fig:apparatus}. During each experimental run, a sessile droplet with a diameter of 1-6 mm is first generated by a syringe pump (Longer Precision Pump) and placed on a super-hydrophobic surface, yielding a contact angle of $\sim140^{\circ}$ and a near-spherical upper part as depicted in Fig.~\ref{fig:setup}(a). Then, the syringe pump generates a smaller-size pendant droplet (1-2 mm in diameter), which is attached to the needle tip. In this experiment, the droplet size ratio $\Delta$ is varied between 1.0 and 5.0. Subsequently, the merging process is initiated by actuating an automatic lifting platform (Winner Optics) which holds the super-hydrophobic surface, slowly bringing the sessile droplet into contact with the pendant droplet. The platform rises at a speed about 10 $\mu$m/s, which is sufficiently small to ignore the gas-film flow disturbing the initial droplet coalescence \cite{CaseSC:08a,CaseSC:09a,PaulsenJD:11a}. 
\begin{figure}[ht!]
\begin{center}
\includegraphics[scale=0.42]{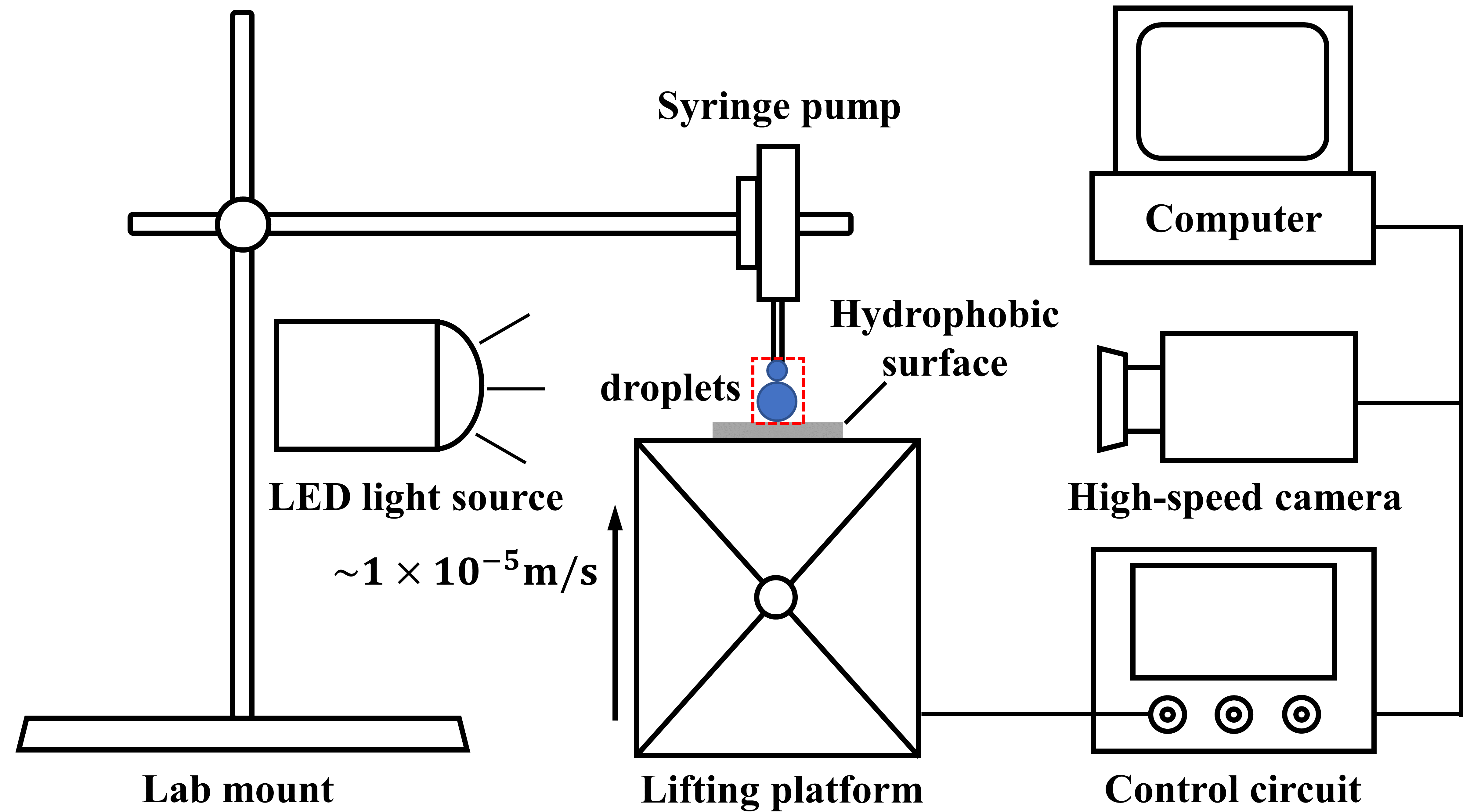}
\vspace{0mm}
\caption{Schematic of the experimental setup for droplet coalescence.}
\label{fig:apparatus}
\vspace{0mm}
\end{center}
\end{figure}

A high-speed camera (Photron SA-Z) integrated with a long-distance microscope (Questar QM100) is used to capture time-resolved shadowgraph images of the merging droplets. The camera operates at 150,000 frame per second (fps) with a spatial resolution of 384$\times$256 pixels and a field of view of 2.04$\times$1.36 mm$^2$, corresponding to a resolution of 5.3 $\mu$m/pixel. The initial time for the coalescence onset is defined based on the first frame showing apparent droplet contact and formation of liquid bridge. As the shutter speed is set at $1/197647$ s, this yields an uncertainty of $\pm 2.53 \times 10^{-6}$ s in measuring the coalescence time. The droplet diameter is determined at the initial state prior to the coalescence, based on fitting an arc to three arbitrarily-selected points on each droplet contour. The uncertainty associated with the diameter measurement is estimated to be within $\pm 3\%$. 

This experiment involves three different liquids, water and two aqueous glycerol solutions with 40 wt$\%$ and 60 wt$\%$ glycerol, respectively. These liquids correspond to $\rho =$ 1000, 1100, and 1150 kg$\cdot$m$^{-3}$, $\mu =$ 1.002, 3.630, and 10.80 mPa$\cdot$s, and $\sigma =$ 72.8, 70.0, and 66.0 mN$\cdot$m$^{-1}$, respectively. The Ohnesorge number $Oh$ varies in the range of $10^{-3}$-$10^{-2}$. The test parameters for all cases are characterized in terms of $\Delta$ and $Oh$, as listed in Table~\ref{tab:S1}. 
\begin{table}
\centering
\caption{Parameters ($D_L$, $D_S$, $\Delta$, and $Oh$) for all experimental cases, sorted in ascending order of $\Delta$ for each fluid type.}
\begin{tabular}{lrrrr}
Fluid type & $D_L$ & $D_S$ & $\Delta$ & $Oh$  \\ \vspace{1.5mm}
           & (mm)  & (mm)  &          &       \\
\midrule
water  & 1.98 & 1.90  & 1.0  & 0.0027 \\
water  & 2.20 & 1.93  & 1.1  & 0.0027 \\
water  & 2.84 & 1.76  & 1.6  & 0.0028 \\
water  & 4.90 & 1.79  & 2.7  & 0.0028 \\
water  & 5.60 & 1.72  & 3.3  & 0.0028 \\
water  & 4.31 & 0.87  & 5.0  & 0.0040 \\
\midrule
\multicolumn{2}{l}{(glycerol solution)} & & & \\
40 wt$\%$  & 3.31 & 2.16 & 1.5 & 0.0089 \\
40 wt$\%$  & 6.11 & 2.15 & 2.8 & 0.0089 \\
40 wt$\%$  & 6.47 & 1.97 & 3.3 & 0.0093 \\
\midrule
\multicolumn{2}{l}{(glycerol solution)} & & & \\
60 wt$\%$  & 3.09 & 1.97 & 1.6 & 0.0279 \\
60 wt$\%$  & 4.75 & 1.81 & 2.6 & 0.0291 \\
60 wt$\%$  & 3.38 & 1.07 & 3.2 & 0.0379 \\
\bottomrule
\end{tabular}
\label{tab:S1}
\end{table}

\subsection*{Experimental data}
The representative image sequences corresponding to the experimental cases plotted in Figs.~\ref{fig:4},~\ref{fig:5}, and~\ref{fig:6} are provided in Fig.~\ref{fig:allimage}. Owing to the limitations in the temporal and spatial resolutions, the data used for the analysis of all cases start from a fixed time of $t = 10^{-4}$ s (corresponding to the 16th frame from the coalescence onset) so as to minimize the undesired measurement uncertainties at the beginning. Therefore, the very early stage of coalescence is not captured or considered in this study. To evaluate how the uncertainties affect the accuracy of our model, we plot the error bars for the data in the $R^{*2}-t^*$ diagram in Fig.~\ref{fig:9}; only the first data point of each case is shown as its error bars are more significant than the rest in the logarithmic coordinate. Comparing different cases, we find that the uncertainties are rather sensitive to the size ratio $\Delta$, as a smaller $\Delta$ generally corresponds to longer error bars. The results again imply a promising agreement between experiment and theory, considering the measurement uncertainties. 
\begin{figure}[ht!]
\begin{center}
\includegraphics[scale=0.37]{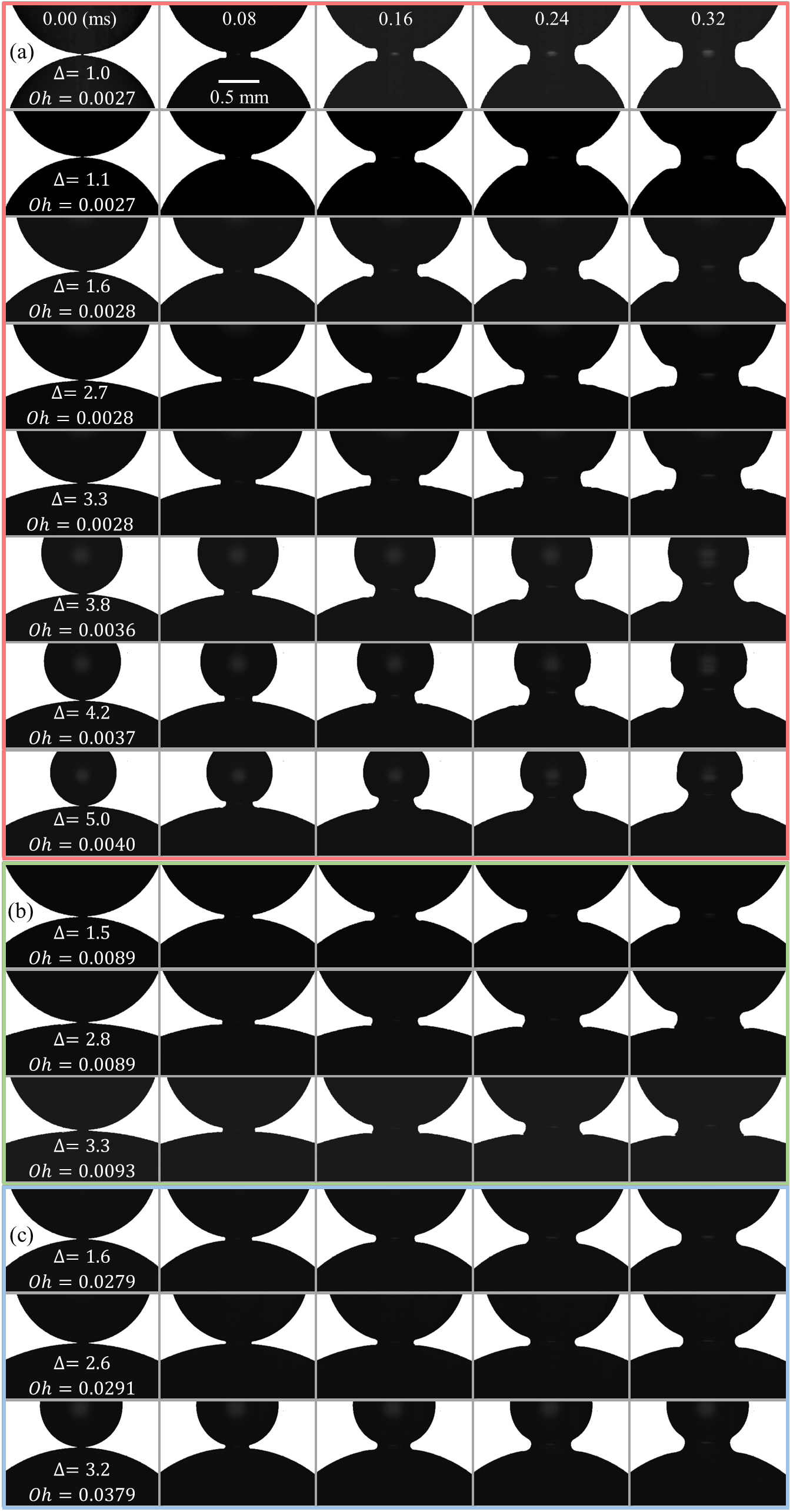}
\vspace{0mm}
\caption{Coalescence image sequences of unequal-size droplets of (a) water, (b) 40wt$\%$, and (c) 60wt$\%$ glycerol solution.}
\label{fig:allimage}
\vspace{0mm}
\end{center}
\end{figure}

\begin{figure}
\begin{center}
\includegraphics[scale=0.44]{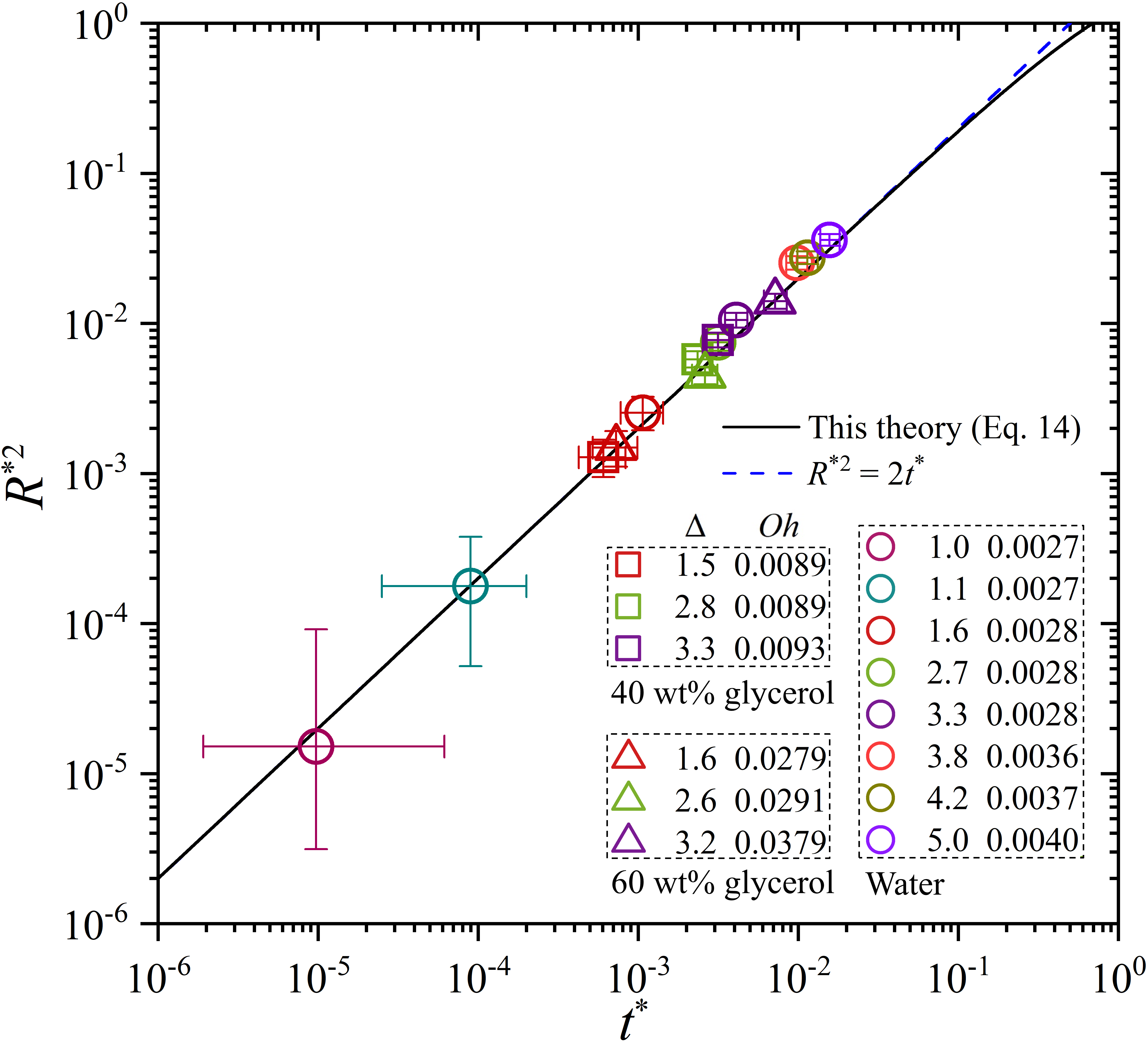}
\vspace{0mm}
\caption{Error bars for the first data point of each case in Fig.~\ref{fig:6}.}
\label{fig:9}
\vspace{0mm}
\end{center}
\end{figure}

\subsection*{Data sharing}
All data obtained experimentally in this work are available for sharing from the corresponding author upon reasonable request.
}

\showmatmethods{} 

\acknow{We acknowledge financial support from the National Natural Science Foundation of China (No. 52176134 and 12072194), the Research Grants Council of the Hong Kong Special Administrative Region, China (Project No. CityU 15218820), and the APRC-CityU New Research Initiatives/Infrastructure Support from Central of City University of Hong Kong (No. 9610601). We thank Dr. Tao Yang and Jiayue Han for their assistance in data processing.}
 
\showacknow{} 

\bibliography{arXiv}

\newcommand{\AIAAJ}{AIAA J.} \newcommand{\AIAAP}{AIAA Paper}
  \newcommand{\ARMA}{Archive for Rational Mechanics and Analysis}
  \newcommand{\ASMEJFE}{J. Fluids Eng., Trans. ASME} \newcommand{\ASR}{Applied
  Scientific Research} \newcommand{\CF}{Computers Fluids}
  \newcommand{\CJFAS}{Can. J. Fish. Aquat. Sci.}
  \newcommand{\ETFS}{Experimental Thermal and Fluid Science}
  \newcommand{\EF}{Experiments in Fluids} \newcommand{\FDR}{Fluid Dynamics
  Research} \newcommand{\IJHMT}{Int. J. Heat Mass Transfer}
  \newcommand{\JASA}{J. Acoust. Soc. Am.} \newcommand{\JCP}{J. Comp. Physics}
  \newcommand{\JEB}{J. Exp. Biol.} \newcommand{\JFM}{J. Fluid Mech.}
  \newcommand{\JMP}{J. Math. Phys.} \newcommand{\JSC}{J. Scientific Computing}
  \newcommand{\JSP}{J. Stat. Phys.} \newcommand{\JSV}{J. of Sound and
  Vibration} \newcommand{\MC}{Mathematics of Computation}
  \newcommand{\MWR}{Monthly Weather Review} \newcommand{\PAS}{Prog. in
  Aerospace. Sci.} \newcommand{\PCPS}{Proc. Camb. Phil. Soc.}
  \newcommand{\PD}{Physica D} \newcommand{\PRA}{Physical Rev. A}
  \newcommand{\PRE}{Physical Rev. E} \newcommand{\PRL}{Phys. Rev. Lett.}
  \newcommand{\PF}{Phys. Fluids} \newcommand{\PFA}{Phys. Fluids A.}
  \newcommand{\PL}{Phys. Lett.} \newcommand{\PRSLA}{Proc. R. Soc. Lond. A}
  \newcommand{\SIAMJMA}{SIAM J. Math. Anal.} \newcommand{\SIAMJNA}{SIAM J.
  Numer. Anal.} \newcommand{\SIAMJSC}{SIAM J. Sci. Comput.}
  \newcommand{\SIAMJSSC}{SIAM J. Sci. Stat. Comput.}
  \newcommand{\TCFD}{Theoret. Comput. Fluid Dynamics} \newcommand{\ZAMM}{ZAMM}
  \newcommand{\ZAMP}{ZAMP} \newcommand{\ICASER}{ICASE Rep. No.}
  \newcommand{\NASACR}{NASA CR} \newcommand{\NASATM}{NASA TM}
  \newcommand{\NASATP}{NASA TP} \newcommand{\ARFM}{Ann. Rev. Fluid Mech.}
  \newcommand{\WWW}{from {\tt www}.} \newcommand{\CTR}{Center for Turbulence
  Research, Annual Research Briefs} \newcommand{\vonKarman}{von Karman
  Institute for Fluid Dynamics Lecture Series}
\begin{thebibliography}{10}

\bibitem{EggersJ:99a}
Eggers J, Lister JR, Stone HA (1999) Coalescence of liquid drops.
\newblock {\em J. Fluid Mech.} 401:293--310.

\bibitem{AartsDGAL:05a}
Aarts DGAL, Lekkerkerker HNW, Guo H, Wegdam GH, Bonn D (2005) Hydrodynamics of
  droplet coalescence.
\newblock {\em Phys. Rev. Lett.} 95(16):164503.

\bibitem{YarinAL:06a}
Yarin AL (2006) Drop impact dynamics: Splashing, spreading, receding,
  bouncing...
\newblock {\em \ARFM} 38:159--192.

\bibitem{ThoravalMJ:12a}
Thoraval MJ, et~al. (2012) von karman vortex street within an impacting drop.
\newblock {\em Phys. Rev. Lett.} 108(26):264506.

\bibitem{TranT:13a}
Tran T, de~Maleprade H, Sun C, Lohse D (2013) Air entrainment during impact of
  droplets on liquid surfaces.
\newblock {\em J. Fluid Mech.} 726:R3.

\bibitem{Thoroddsen:05a}
Thoroddsen ST, Takehara K, Etoh TG (2005) The coalescence speed of a pendent
  and a sessile drop.
\newblock {\em J. Fluid Mech.} 527:85--114.

\bibitem{PaulsenJD:11a}
Paulsen JD, Burton JC, Nagel SR (2011) Viscous to inertial crossover in liquid
  drop coalescence.
\newblock {\em Phys. Rev. Lett.} 106(11):114501.

\bibitem{ZhangP:11a}
Zhang P, Law CK (2011) An analysis of head-on droplet collision with large
  deformation in gaseous medium.
\newblock {\em Phys. Fluids} 23(4):042102.

\bibitem{KavehpourHP:15a}
Kavehpour HP (2015) Coalescence of drops.
\newblock {\em Annu. Rev. Fluid Mech.} 47:245--268.

\bibitem{EggersJ:25a}
Eggers J, Sprittles JE, Snoeijer JH (2025) Coalescence dynamics.
\newblock {\em Annu. Rev. Fluid Mech.} 57:61--87.

\bibitem{DucheminL:03a}
Duchemin L, Eggers J, Josseran C (2003) Inviscid coalescence of drops.
\newblock {\em J. Fluid Mech.} 487:167--178.

\bibitem{WuM:04a}
Wu M, Cubaud T, Ho CM (2004) Scaling law in liquid drop coalescence driven by
  surface tension.
\newblock {\em Phys. Fluids} 16(7):L51--L54.

\bibitem{FezzaaK:08a}
Fezzaa K, Wang Y (2008) Ultrafast x-ray phase-contrast imaging of the initial
  coalescence phase of two water droplets.
\newblock {\em Phys. Rev. Lett.} 100(10):104501.

\bibitem{CaseSC:09a}
Case SC (2009) Coalescence of low-viscosity fluids in air.
\newblock {\em Phys. Rev. E} 79(2):026307.

\bibitem{BurtonJC:07a}
Burton JC, Taborek P (2007) Role of dimensionality and axisymmetry in fluid
  pinch-off and coalescence.
\newblock {\em Phys. Rev. Lett.} 98(22):224502.

\bibitem{PothierJC:12a}
Pothier JC, Lewis LJ (2012) Molecular-dynamics study of the viscous to inertial
  crossover in nanodroplet coalescence.
\newblock {\em Phys. Rev. B} 85(11):115447.

\bibitem{GrossM:13a}
Gross M, Steinbach I, Raabe D, Varnik F (2013) Viscous coalescence of droplets:
  A lattice boltzmann study.
\newblock {\em Phys. Fluids} 25(5):052101.

\bibitem{SprittlesJE:12a}
Sprittles JE, Shikhmurzaev YD (2012) Coalescence of liquid drops: Different
  models versus experiment.
\newblock {\em Phys. Fluids} 24(12):122105.

\bibitem{YaoW:05a}
Yao W, Maris HJ, Pennington P, Seidel GM (2005) Coalescence of viscous liquid
  drops.
\newblock {\em Phys. Rev. E} 71(1):016309.

\bibitem{PaulsenJD:13a}
Paulsen JD (2013) Approach and coalescence of liquid drops in air.
\newblock {\em Phys. Rev. E} 88(6):063010.

\bibitem{Xia:19b}
Xia X, He C, Zhang P (2019) Universality in the viscous-to-inertial coalescence
  of liquid droplets.
\newblock {\em Proc. Natl. Acad. Sci. U.S.A.} 116(47):23467--23472.

\bibitem{HackMA:20a}
Hack MA, et~al. (2020) Self-similar liquid lens coalescence.
\newblock {\em Phys. Rev. Lett.} 124:194502.

\bibitem{AnilkumarAV:91a}
Anilkumar AV, Lee CP, Wang TG (1991) Surface-tension-induced mixing following
  coalescence of initially stationary drops.
\newblock {\em Phys. Fluids A} 3:2587.

\bibitem{BlanchetteF:10a}
Blanchette F (2010) Simulation of mixing within drops due to surface tension
  variations.
\newblock {\em Phys. Rev. Lett.} 105:074501.

\bibitem{ZhangP:13a}
Liu D, Zhang P, Law CK, Guo Y (2013) Collision dynamics and mixing of
  unequal-size droplets.
\newblock {\em Int. J. Heat Mass Transfer} 57:421.

\bibitem{ZhangP:15a}
Sun K, Wang T, Zhang P, Law CK (2015) Non-newtonian flow effects on the
  coalescence and mixing of initially stationary droplets of shear-thinning
  fluids.
\newblock {\em Phys. Rev. E} 91:023009.

\bibitem{ZhangP:16a}
Tang C, Zhao J, Zhang P, Law CK, Huang Z (2016) Dynamics of internal jets in
  the merging of two droplets of unequal sizes.
\newblock {\em J. Fluid Mech.} 795:671--689.

\bibitem{Xia:17a}
Xia X, He C, Yu D, Zhao J, Zhang P (2017) Vortex-ring-induced internal mixing
  upon the coalescence of initially stationary droplets.
\newblock {\em Phys. Rev. Fluids} 2(11):113607.

\bibitem{SM:24a}
{See Supplemental Materials for a parameter list of all experimental cases.}
  (year?).

\bibitem{CaseSC:08a}
Case SC, Nagel SR (2008) Coalescence in low-viscosity liquids.
\newblock {\em Phys. Rev. Lett.} 100(8):084503.

\end{thebibliography}

\end{document}